\documentclass[conference]{IEEEtran}
\IEEEoverridecommandlockouts
\usepackage{amsmath}
\usepackage{array}
\usepackage{amsthm}
\usepackage{enumitem}
\usepackage{graphicx, subfigure}
\usepackage{amssymb}
\usepackage{cite}
\usepackage{url}
\usepackage[ruled,vlined,lined,ruled,linesnumbered]{algorithm2e}
\usepackage{relsize}
\usepackage[nodisplayskipstretch]{setspace}
\usepackage{longtable}
\usepackage{booktabs}
\usepackage{multirow}
\usepackage{soul} 
\usepackage[subnum]{cases}
\usepackage{color}

\setstcolor{red}

\title{Machine Learning for Heterogeneous Ultra-Dense Networks \textcolor{black}{with Graphical Representations}}
\author{Congmin Fan, Ying-Jun Angela Zhang, and Xiaojun Yuan}
\begin{document}

\maketitle
\begin{abstract}
Heterogeneous ultra-dense network (H-UDN) is envisioned as a promising solution to sustain the explosive mobile traffic demand through network densification. By placing access points, processors, and storage units as close as possible to mobile users, H-UDNs bring forth a number of advantages, including high spectral efficiency, high energy efficiency, and low latency. Nonetheless, the high density and diversity of network entities in H-UDNs introduce formidable design challenges in collaborative signal processing and resource management. This article illustrates the great potential of machine learning techniques in solving these challenges. In particular, we show how to utilize graphical representations of H-UDNs to design efficient machine learning algorithms. 
\end{abstract}

\section{Introduction}
Driven by the astounding development of smart phones, mobile applications and the Internet of Things (IoT), traffic demand grows exponentially in mobile networks. Heterogeneous ultra-dense networks (H-UDNs) (such as femtocell/picocell, cloud radio access network (Cloud-RAN), and fog radio access network (Fog-RAN)) have emerged as a promising solution to sustain the enormous mobile traffic demand. The coverage and capacity of wireless networks are improved through densified access points (APs) and communication links, which greatly enhances the spatial reuse of limited frequency resources. 
\par 
H-UDNs, however, bring opportunities as well as formidable challenges. A H-UDN is expected to support an abundance of new applications with various new service requirements. For instance, massive machine-type communications require high connection density; auto pilot cars require low latency and ultra-high reliability; augmented reality requires both high throughput and low latency. To effectively and efficiently sustain these manifold requirements in a complicated system like a H-UDN is undoubtedly a mission worth pursuing. In addition, the existing advanced techniques, such as multi-cell coordination and massive multiple-input multiple-output (MIMO), cannot be easily extended to a H-UDN. This is because many assumptions used in the existing techniques are no longer valid or accurate in H-UDNs. For example, the law of large numbers and the random matrix theory have been widely used to simplify the analysis of massive MIMO. \textcolor{black}{However, due to random scattering of APs and users, the channels between APs and users follow heavy-tailed distributions that are not analyzable for algorithm design using the existing random matrix theory \cite{fan2017scalable}.} Moreover, the high density of devices leads to prohibitively high complexity if the existing algorithms are directly applied. Consider, for example, a system supporting thousands or even tens of thousands of machine-type devices in a small area. The complexity of jointly detecting the signals sent by the devices using, say, a simple linear minimum mean square error (LMMSE) detector would cause unaffordable computational complexity, since the complexity of LMMSE grows cubically in the number of terminals.
\par
\textcolor{black}{Machine learning is a family of promising techniques to address the above-mentioned challenges. Unlike the traditional model-based approaches that are optimized based on mathematically convenient models, machine-learning based approaches are driven by real-world data, and thus are less sensitive to model imperfections. As such, machine learning provides means of system optimization through learning from data, and has strong potential to improve system performance in reliability, generality, adaptivity, and efficiency.} Previously, machine learning has been extensively used to solve a wide variety of problems in image/audio processing, social behavior analysis, project management, etc. The applications of machine learning in wireless networks start to attract research interests in recent years. As discussed in \cite{jiang2017machine}, \cite{wang2016survey} and the references therein, machine learning approaches have potential applications in cognitive radios, massive MIMOs, device-to-device communications, etc. The goal of this paper is to complement their contributions by investigating the use of machine learning in H-UDNs, especially for solving collaborative signal processing and resource allocation problems. Specifically, we discuss how to utilize graphical representations of H-UDNs to design efficient machine learning algorithms. We first introduce several recently proposed signal processing algorithms (namely, randomized message passing \cite{fan2017scalable}, bilinear generalized approximate message passing (BIG-AMP) \cite{zhang2017blind}, and deep learning \cite{borgerding2016onsager}) based on the coverage graph of APs. \textcolor{black}{Then, we extend the discussion to resource management problems, such as radio resource allocation, power allocation, and cache placement.} Reinforcement learning, deep learning, and semi-supervised learning are introduced as potential solutions, where the graphical models based on certain features of H-UDNs can help greatly improve the efficiency of the solutions.

\section{H-UDNs and Graphical Representations}
\begin{figure*}
    \centering
    {\includegraphics[width=0.8\textwidth]{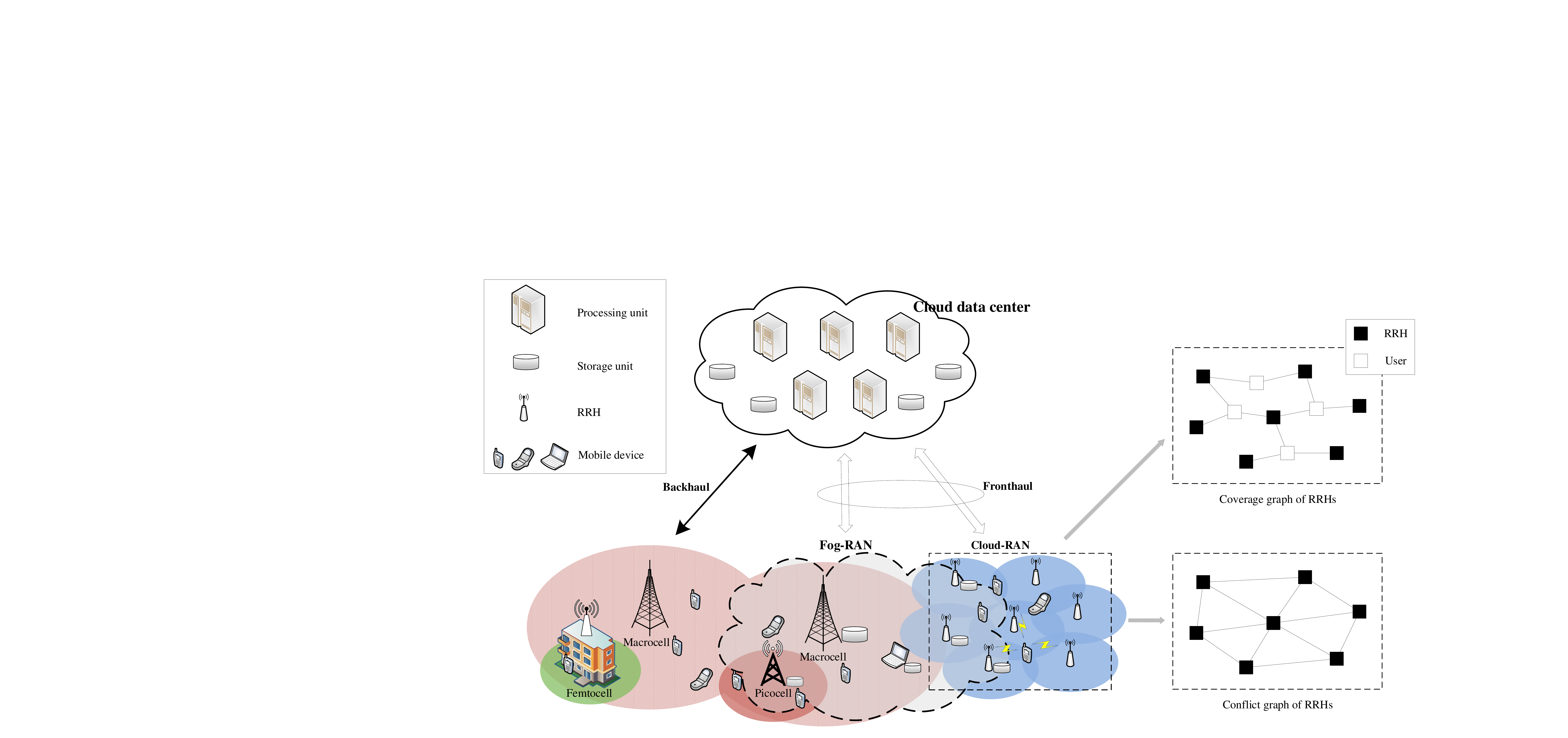}}
    \caption{A H-UDN consists of a femtocell, a picocell, a macrocell, a Cloud-RAN, and a Fog-RAN, where the Fog-RAN incorporates the macrocell and part of the Cloud-RAN. A coverage graph and a conflict graph are induced from the Cloud-RAN.}
    \label{fig:my_label}
\end{figure*}
In this section, we first introduce the main entities of a H-UDN. Then, H-UDNs are modelled as graphs that depict various interactions between network entities. Based on these graphs, machine learning algorithms will be discussed in later sections for efficient network operations.
\subsection{H-UDNs}
A H-UDN consists of various types of network architectures; see Fig. 1. In the following, we introduce typical network architectures as well as their unique features.
\subsubsection{Macrocell and Picocell/Femtocell} 
Macrocells are cells that provide radio coverage served by high power base stations (BSs) in a cellular network. Picocells and femtocells are served by small and low-power BSs to provide uninterrupted coverage for end users even in areas difficult or expensive to cover by macrocells. In particular, femtocells are deployed, powered, and connected by end users or small businesses with less active control from network operators. Hence, the operation of femtocells is more autonomous than picocells and macrocells. As such, multi-cell coordination in a H-UDN must consider different levels of processing power, backhaul limitation, information availability, controllability, and willingness of participation of different cells.
\subsubsection{Cloud-RAN}
A Cloud-RAN consists of three key components: (i) the distributed remote radio heads (RRHs), (ii) a pool of baseband processing units (BBUs) in a data center cloud and (iii) a high-bandwidth and low-latency optical transport network connecting BBUs and RRHs. Compared with traditional BSs, the RRHs are lightweight, allowing them to be deployed in a high density with low cost. Meanwhile, the centralized BBU pool enables seamless coordination of RRHs for collaborative signal processing, radio resource allocation, network virtualization, etc. Thus, a natural challenge is to design scalable RRH coordination algorithms to avoid high overhead and computational cost caused by the high density of RRHs.

\subsubsection{Fog-RAN}
The key idea of Fog-RAN is to take full advantages of local computing, communication, and storage capabilities at edge devices (such as RRHs, smartphones, laptops, etc.), so as to avoid heavy communication overhead and large latency caused by backhaul/fronthaul transmission and centralized processing. To fully utilize the capabilities at edge devices, tasks should be efficiently decomposed and assigned to different devices. As such, decentralized control is critical in Fog-RAN. As a task offloading scheme, Fog-RAN is usually overlaid on other network architectures. For example, the Fog-RAN in Fig. 1 partially merges with a macrocell and a Cloud-RAN.

\par 
The above-mentioned network architectures may coexist in a single H-UDN. For example, coordination among devices may rely on centralized processing in the BBU pool of a Cloud-RAN. Meanwhile, the computational tasks may be offloaded to network edges of a Fog-RAN for delay-sensitive applications. In a nutshell, a H-UDN is an inseparable continuum of different types of networks. Signal processing and resource management in such a complicated system become challenging due to the close interaction between different network entities. As a result, interference across different networks must be carefully managed. Likewise, different types of resource allocation, including allocation of radio, computation, and storage, are to be coordinated at different levels of network hierarchies and different types of network entities. To fully utilize the resources of network entities without causing significant implementational and computational costs, it is of utmost importance to design efficient resource management schemes for H-UDNs.

\subsection{Graphical Representations}\label{S:GR}
In this subsection, we describe the interactions between different entities of a H-UDN using graphical models.
\begin{figure*}[!t]
\subfigure[A factor graph based on the equation $P_{XYZ}(x,y,z)=P_X(x)P_{Y|X}(y|x)P_{Z|Y}(z|y)$.]{\includegraphics[width=0.45\textwidth]{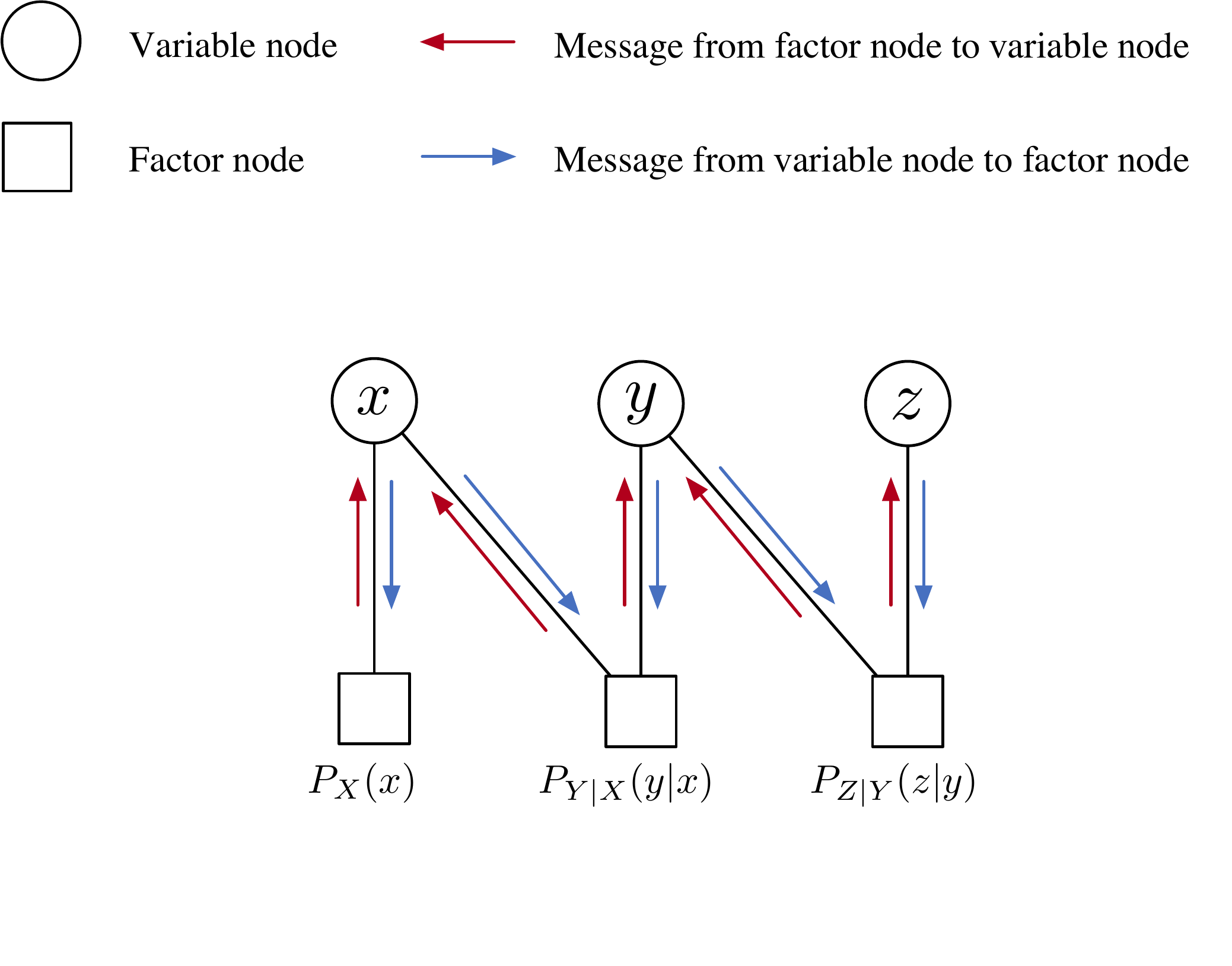}}
\hspace{0.3em}
\subfigure[A factor graph based on Equation (2), where $x_k$ is the signal transmitted by user $k$, $y_n = \sum_{k=1}^Kh_{n,k}x_k+n_n$ is the signal received by AP $n$, $h_{n,k}$ is the channel coefficient from user $k$ to AP $n$, and $n_n$ is the noise at AP $n$.]{\includegraphics[width=0.55\textwidth]{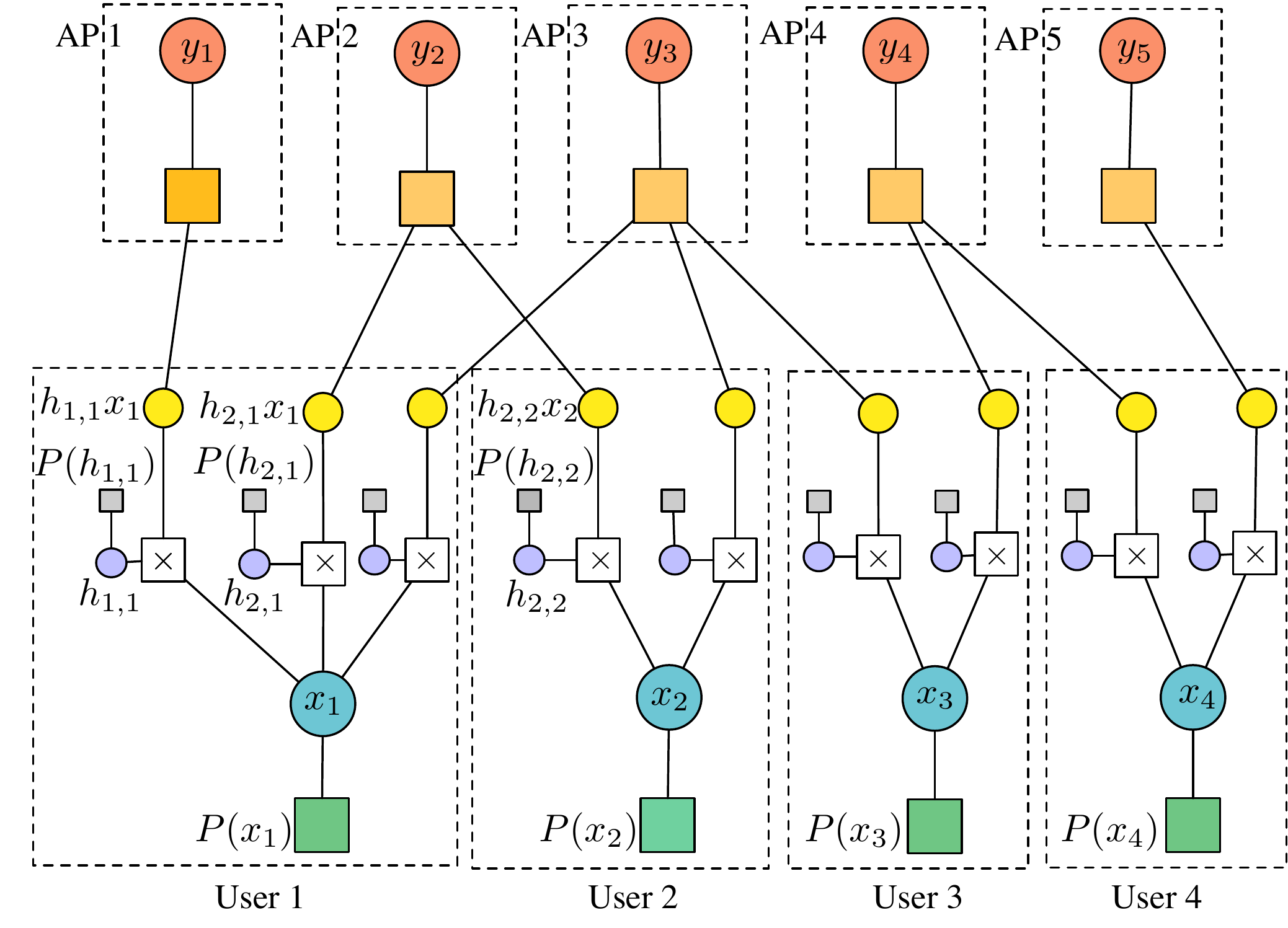}}
\caption{Two examples of factor graph}
\end{figure*}
\subsubsection{Coverage}
The coverage of APs in a H-UDN can be modelled as a bipartite graph as shown in Fig. 1, where the APs (denoted by solid squares) and users (denoted by hollow squares) are treated as vertices. An edge connects an AP and a user if the AP serves the user. Similarly, we can extend this graphical model to the coverage of caches, computing units, etc.
\subsubsection{Conflict} 
Conflicts exist in a H-UDN due to limited resources. For example, nearby links interfere with each other, and thus cannot transmit on the same frequency channel at the same time. Likewise, multiple BSs compete for limited fronthaul and backhaul capacity. Different tasks compete for storage and computing power. By connecting the conflicting entities with edges, a conflict graph is constructed.

\par 
Besides the discussions above, H-UDNs can be represented by graphs in many other aspects, such as cooperation among network entities, dependency between content popularity and caching placement, time-varing connection between users and APs, interaction between different network levels, etc. Most of these graphs share common properties. First, most of them are sparse, i.e., a limited number of edges exist between vertices. Second, most of the graphs are locally dense. That is, the graphs contain many short loops. These two properties are due to the scattered physical locations of network entities. Nearby entities are more likely to interfere with each other. It is also easier for them to share information and cooperate. Meanwhile, interactions between faraway entities are weak and usually negligible. In what follows, we will discuss how to exploit these two properties to design efficient machine learning algorithms for signal processing and resource management in H-UDNs.

\section{Signal Processing in H-UDNs}
In a UDN, interference is significantly magnified due to the high density of APs. To fully exploit the benefits of network densification, efficient interference management schemes become essential. It is commonly believed that network cooperation is an effective technique to mitigate interference and enhance network performance. However, cooperative signal processing in UDNs is usually very costly due to the following reasons. First, the amount of data in a H-UDN is extremely large. Jointly processing these data may involve prohibitively high computational complexity. Secondly, full-scale cooperation requires the estimation of a sufficiently large channel matrix that consists of channel coefficients from all mobile users to all APs. This causes significant channel estimation overhead, and thereby fundamentally limits the cooperation gain. Thirdly, with limited capacity of the fronthual network, excessive information sharing among APs is impossible. In this section, we will discuss potential solutions to the above issues based on the graphical representations of the H-UDN and the corresponding machine learning techniques. 

\subsection{Signal Processing Based on Factor Graph}
We first illustrate the use of factor graphs and the associated message passing (MP) algorithms \cite{loeliger2007factor} for efficient signal processing in H-UDNs. A factor graph is a bipartite graph representing the factorization of a function, especially a probabilistic function. Assume, for example, that $X, Y,$ and $Z$ are random variables that form a Markov chain. Then, their joint probability mass function $P_{XYZ}(x,y,z)$ can be factored as a product of $P_X(x)$, $P_{Y|X}(y|x)$, and $P_{Z|Y}(z|y)$. The factorization can be
expressed by a factor graph shown in Fig. 2(a). Specifically, the factor graph consists of variable nodes $\{x,y,z\}$ (denoted by circles), factor nodes $\{P_X, P_{Y|X}, P_{Z|Y}\}$ (denoted by squares), and edges. A variable node and a factor node are connected by an edge only when the variable is an input of the corresponding factor function. That is, edges specify the dependence relations between two types of nodes. 
\par 
A typical statistical inference problem is to infer unobserved variables based on observed variables and a given probability model. With the help of the factor-graph representation of the probability model, the problem translates to estimating the values of unobserved nodes conditioned on the values of observed ones. Efficient MP algorithms associated with graphical representations can be designed to solve the inference problem. In a MP algorithm, there are two types of messages over an edge:
\begin{itemize}
    \item Messages from a factor node to a variable node.
    \item Messages from a variable node to a factor node.
\end{itemize}
Each message over an edge is designed to capture the probabilistic information from nearby nodes. Each message is iteratively updated only based on the previous messages sent by nearby nodes. In this way, MP provides a distributed solution for the inference problem. Moreover, when the factor graph is sparse, the complexity can be further reduced. Notice that many signal processing problems in communications are inference problems, since the receiver in a communication system wants to estimate the unobserved transmit signals by processing its observed receive signals. Hence, factor-graph-based MP has a great potential to provide an efficient solution for signal processing in H-UDNs. However, depending on specific problems, MP algorithms need to be carefully adjusted to achieve satisfactory performance. In the following, we illustrate how to utilize the special features of the underlying systems to design efficient message-passing algorithms for signal processing. In particular, we will discuss the applications of factor graph in signal detection problems with and without perfect channel state information (CSI).

\subsubsection{Signal Processing with Perfect CSI}
In this example, we consider joint uplink signal detection in a Cloud-RAN. Suppose that the APs (i.e., RRHs) perfectly know the channel matrix $\mathbf H$, and the signals transmitted by the users follow an independent distribution. Then, the widely-used maximum\textit{ a posteriori} (MAP) estimator can be expressed as $\widehat{\mathbf{x}}=\arg \max \prod_{n=1}^N P(y_n|\mathbf{x},\mathbf{H})\prod_{k=1}^KP(x_k)$,
where $y_n$ is the signal received by AP $n$, and $x_k$ is the signal transmitted by user $k$. The corresponding factor graph is given in Fig. 2(b). By treating the dashed rectangles as APs and users, the factor graph can be viewed as a graph induced from the coverage graph. More importantly, the factor graph shares certain common properties with the coverage graph. Recall that the coverage graph is a sparse and locally dense graph since each AP can only receive reasonably strong signals from a small number of nearby users. Hence, the factor graph is also globally sparse and locally dense. On one hand, the high sparsity leads to a significant reduction of computational complexity of messages in the message-passing algorithm. On the other hand, the local denseness compromises the convergence performance of the message-passing algorithm. MP converges when the factor graph is a tree or only contains random long loops. Unfortunately, locally dense factor graphs contain a large number of short loops. To improve the convergence of MP, several efficient methods have been proposed, such as damping and asynchronous updating. With damping, an updated message is a weighted average of the message in the previous iteration and the message calculated by the original message updating rule. With asynchronous updating, the messages are updated sequentially instead of in parallel, with the hope to break the effect of short loops if the updating schedule is appropriately chosen. However, so far there is no efficient and systematic methods to find an updating schedule that ensures convergence of MP. Recently, Ref. \cite{fan2017scalable} proposed a random asynchronous updating strategy to improve the convergence performance, where the messages are updated sequentially in a random order. Fig. 3(a) and Fig. 3(b) compare the convergence performance of MP with other widely-used algorithms. As shown in Fig. 3(a), the MP algorithms converge much faster than the other algorithms. Through Fig. 3(b), we observe that the number of iterations needed by MP does not increase with the network size, while that of the other algorithms grows roughly linear with the network size. This implies that MP with appropriate adjustment is an effective solution for scalable signal processing in large-scale H-UDNs.

\begin{figure}

    \subfigure[Relative error versus the number of iterations for various detection algorithms, including message passing (MP), conjugate gradient (CG), generalized approximate message passing (GAMP), and alternating direction method of multipliers (ADMM), when the number of APs is $40$, the number of users is $32$, the network area is $10\text{km}^2$, and the average
transmit signal-to-noise ratio at the user side is $95$dB]{
    \includegraphics[width=0.45\textwidth]{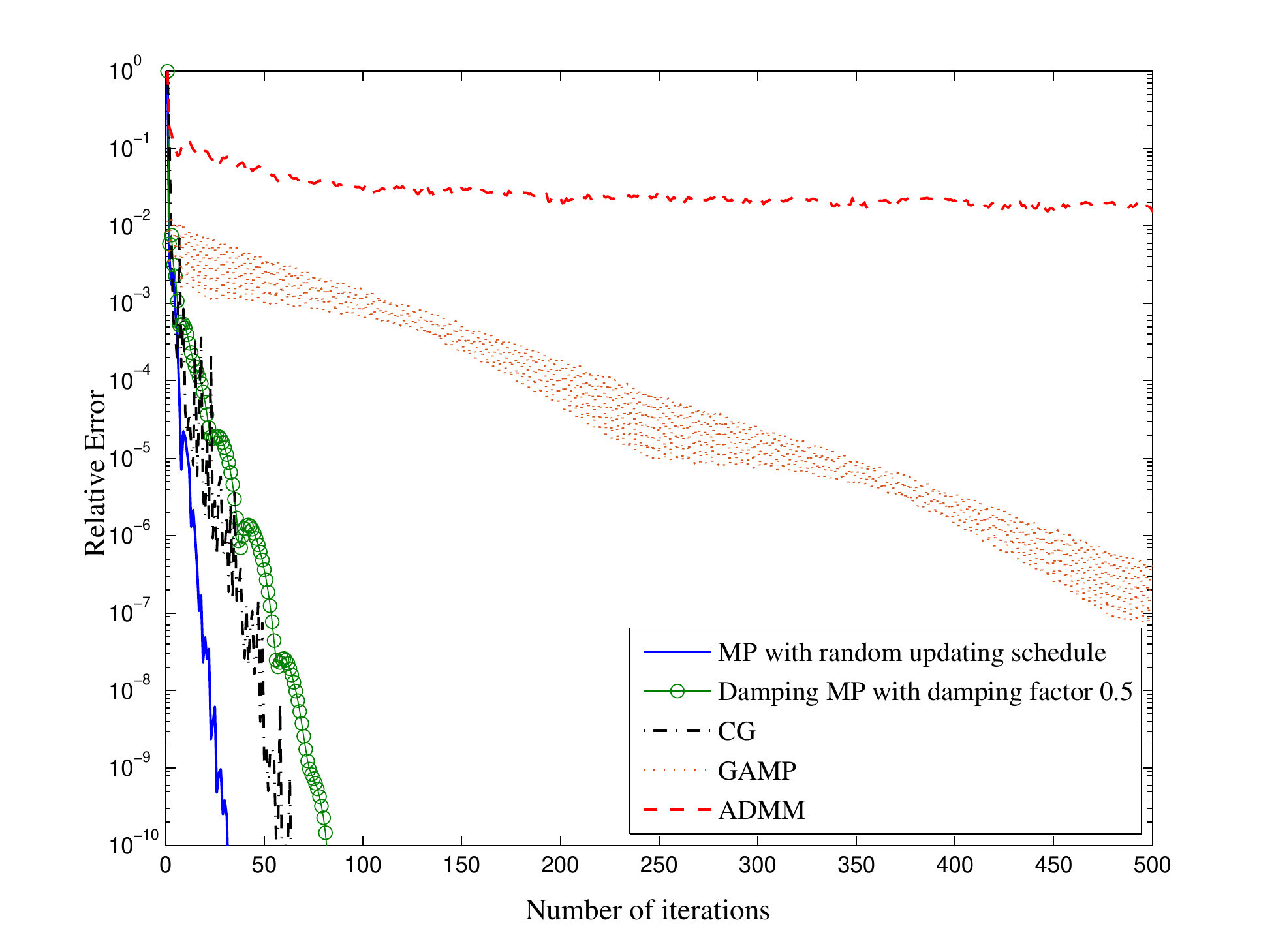}}
    	\subfigure[Convergence speed against the network size for various detection algorithms, including MP, CG, GAMP, when the density of APs is $10/\text{km}^2$, the density of users is $8/\text{km}^2$, and the average
transmit signal-to-noise ratio at the user side is $95$dB.]
	{\includegraphics[width=0.45\textwidth]{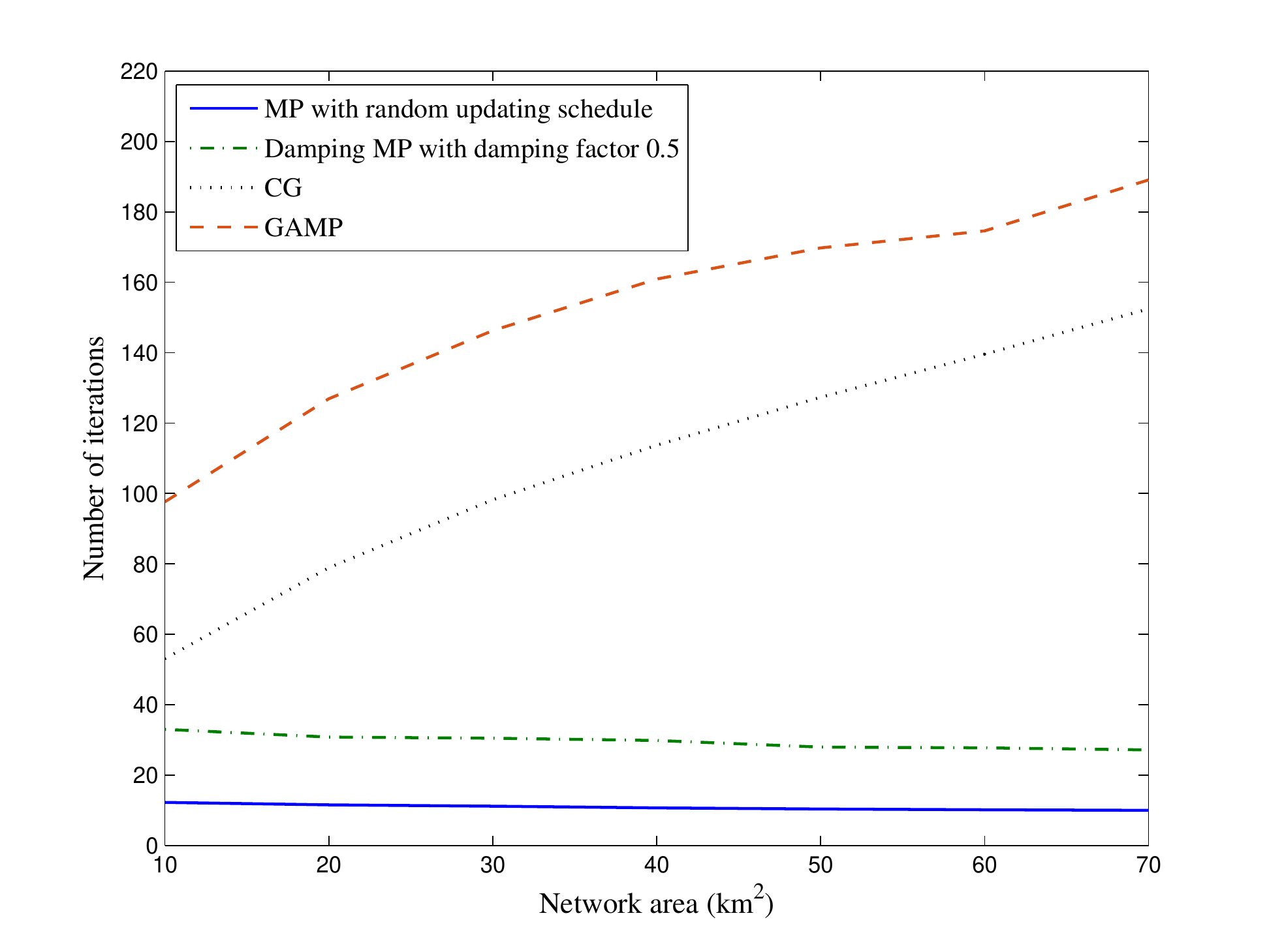}}
\label{fig:prob_con}
\subfigure[Performance comparison among various detection schemes when the number of receive antennas is 500, the number of users is $K = 50$, and the coherence time $T = 100$.]{\includegraphics[width=0.4\textwidth]{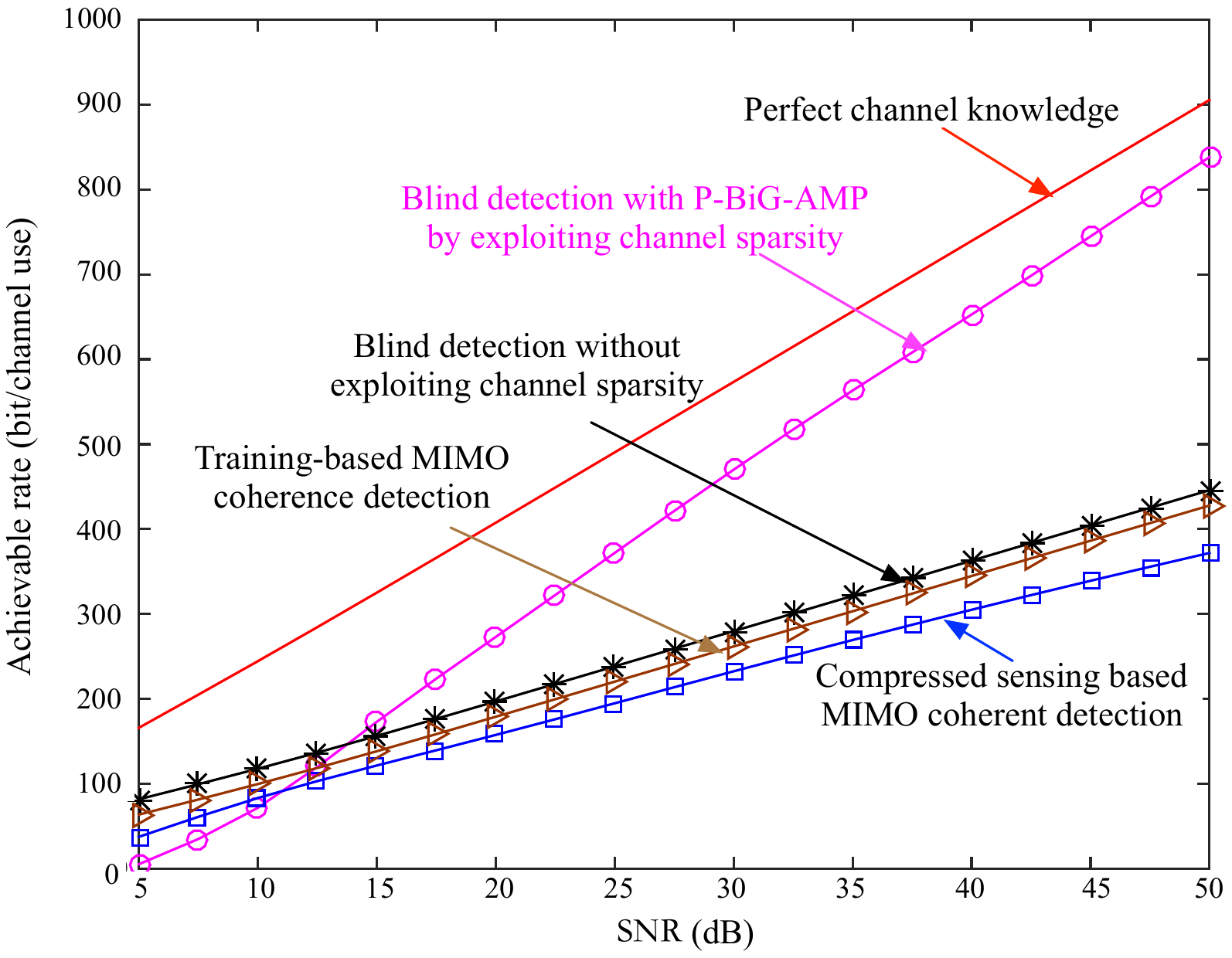}}
    \caption{Convergence speed and performance of message passing (MP).}
\end{figure}

\subsubsection{Blind Signal Detection}
In the previous example, we have assumed perfect CSI at the receiver side. In practical systems, CSI acquisition consumes a substantial amount of system resource when the network size is large, and thus fundamentally limits the system capacity. To avoid the channel estimation overhead, another line of research works on blind-detection approaches, where both the channel matrix $\mathbf{H}$ and the transmitted signals $\mathbf{x}$ are estimated simultaneously from the received signals $\mathbf{y}$. However, it has been proved that blind approaches can, at best, achieve the same degrees of freedom as training-based approaches due to the fact that an unknown channel makes signal detection in a blind approach more difficult. Recently, it was shown that channel sparsity in certain transformed domain can be exploited to improve the performance of blind detection. For example, in massive MIMO systems, the channel matrix $\mathbf{H}$ is sparse in the angular domain. The MAP estimation can be written as $(\widehat{\mathbf{x}},\widehat{\mathbf{H}})=\arg \max_{\mathbf{x},\mathbf{H}} P(\mathbf{y}|\mathbf{x}, \mathbf{H})P(\mathbf{x})P(\mathbf{H})$,
where the sparsity of $\mathbf{H}$ is incorporated in its prior distribution $P(\mathbf{H})$. Similar to the previous example, the above MAP estimation
can also be represented by a factor graph. With careful design, MP algorithms, such as projected bilinear generalized approximate message passing (P-BiG-AMP) \cite{zhang2017blind}, can be potentially applied to solve the blind detection problem. Fig. 3(c) shows that the P-BiG-AMP algorithm significantly outperforms other existing detection schemes, and performs closely to the ideal case with perfect channel knowledge.
\par 
\begin{figure}
    \subfigure[An M-layer neural network based on AMP, with tunable updating matrices $\{\mathbf{B}_1,\mathbf{B}_2,\cdots,\mathbf{B}_M\}$.]{
    \includegraphics[width=0.5\textwidth]{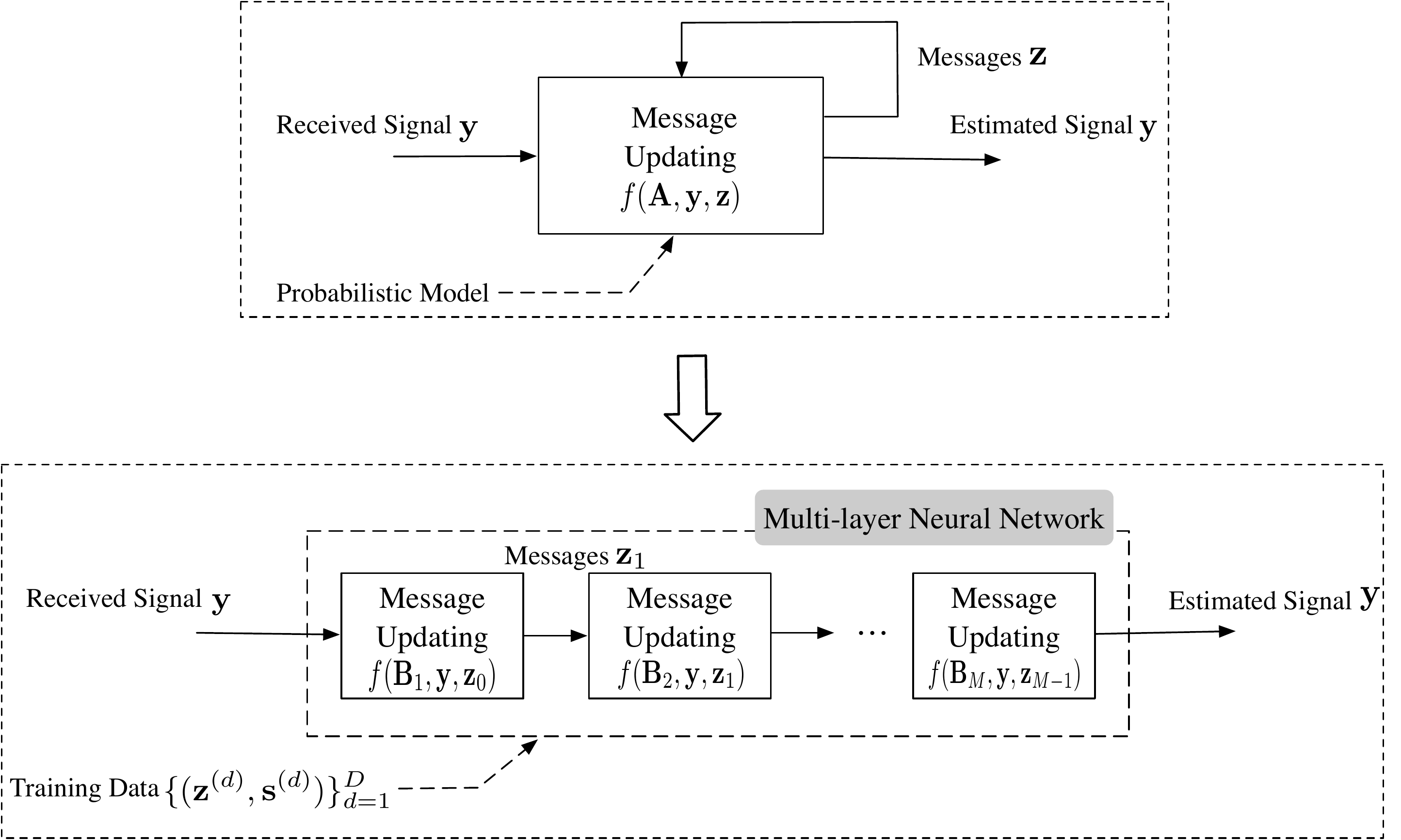}}
    \centering
    	\subfigure[A fully-connected neural network to approximate WMMSE.]{\includegraphics[width=0.48\textwidth]{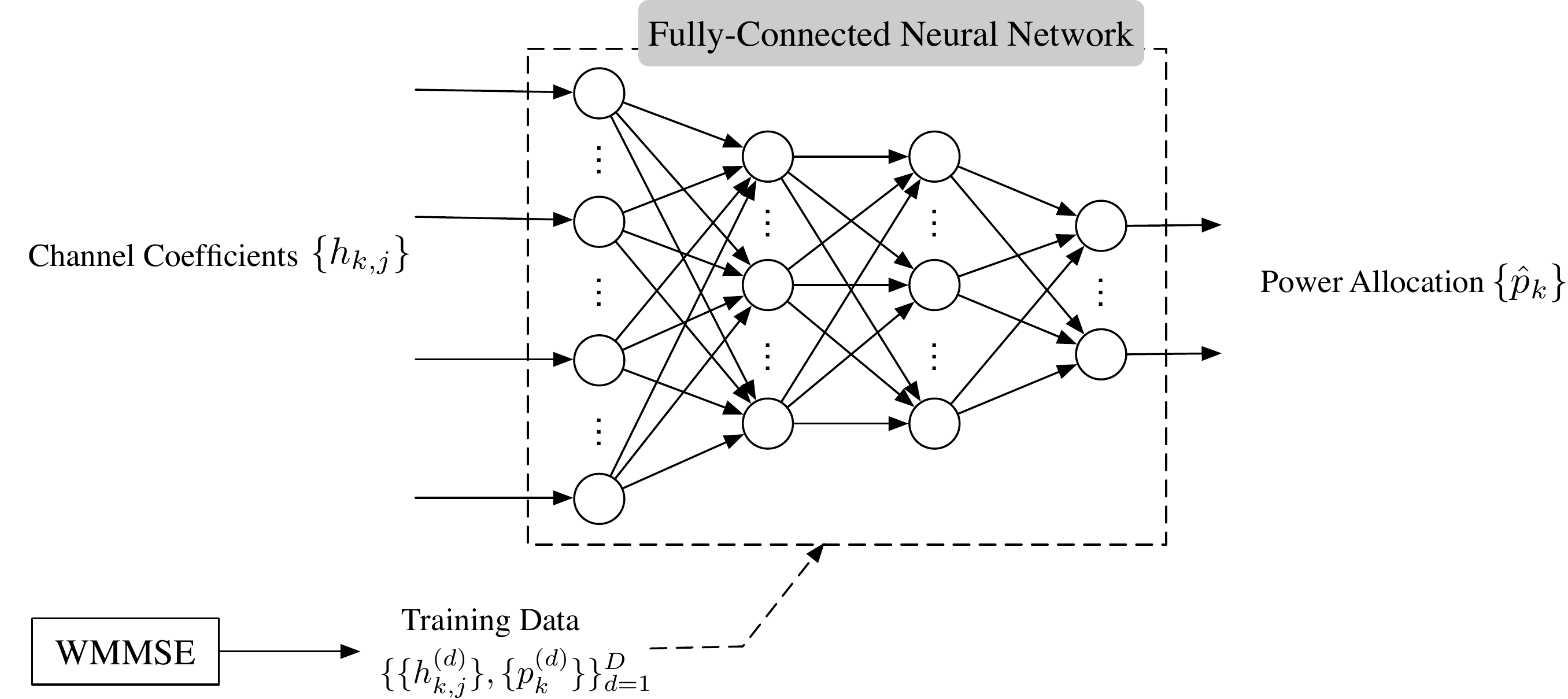}}
    \caption{Neural networks to approximate existing signal processing/resource management algorithms.}
\end{figure}

\subsection{Signal Processing with Sparse Neural Networks}
In this subsection, we discuss the design of efficient deep learning algorithms for signal processing in H-UDNs. In deep learning, a multi-layer neural network is trained to imitate the relationship between the input and output data based on a large training set. Once the neural network is well trained, it can be used to complete the task it was trained for with very low complexity. Thanks to its outstanding performance, e.g. in image classification and speech recognition, deep learning has been considered as a promising tool for future wireless network design. However, how to adapt existing deep learning techniques to wireless systems is still an open problem. Most of the previous work in image/audio processing has used very little prior information during the training of neural networks. Nonetheless, in wireless systems, there exists a large amount of prior information, including user locations, APs' coverage area, distributions of channel gains, and so on. Ignoring such prior information would inevitably cause a performance loss. Recently, noticeable efforts have been made to utilize prior information to improve the performance of deep learning \cite{nachmani2016learning, borgerding2016onsager}. In \cite{borgerding2016onsager}, a novel neural-network architecture has been proposed to mimic the updating rule of a well-known signal processing algorithm, approximate message passing (AMP). As shown in Fig. 4(a), the message updating function in each layer of the neural network is the same as that of AMP, but the updating matrices $\{\mathbf{B_1},\mathbf{B}_2,\cdots,\mathbf{B}_M\}$ are learned from the training data (e.g., a set of received and transmitted signals $\{(\mathbf{y}^{(d)},\mathbf{x}^{(d)})\}_{d=1}^D$) instead of being pre-determined by the channel matrix $\mathbf{A}$. The proposed neural network significantly outperforms the original AMP algorithm in both computational time and accuracy. In addition, Ref. \cite{borgerding2016onsager} shows that fixing the structure of updating matrices according to the channel matrix $\mathbf{A}$, i.e., letting $\mathbf{B}_m=\mathbf{C}_m\mathbf{A}$, renders a more efficient learning procedure. This is because the channel matrix, as prior information, represents the relationship between the channel input and the channel output. When applying to large-scale UDNs, however, the approach in \cite{borgerding2016onsager} might be impractical due to the high computational complexity and signaling overhead to estimate the channel matrix. On the other hand, as discussed in Subsection \ref{S:GR}, the coverage graph of a H-UDN is often sparse and easy to obtain. Thus, it is natural to restrict the updating matrices in the neural networks to sparse matrices according to the coverage information. This implies that the neural network can be constructed as a sparse network, thereby significantly reducing the number of tunable parameters. As a result, the complexity of neural network training can be drastically lowered.

\section{Resource Management in H-UDNs}
To date, most resource allocation problems in wireless networks have been formulated and solved as optimization problems, thanks to the availability of efficient algorithms when the formulation is in a nice (e.g., convex) form. One limitation of optimization methods is that they heavily rely on rigidly defined and mathematically convenient models. However, in H-UDNs, complicated interactions between various network entities render accurate modeling difficult. Even if an accurate model can be constructed, it is likely too tedious to be mathematically tractable. In contrast, driven by real-world data instead of pre-assumed models, machine learning algorithms hold significant potential. In this section, we discuss the design of efficient machine learning algorithms based on graphical models for resource management in H-UDNs; see the summary in Table \ref{table:t1}.

\begin{table*}
\small
	        \caption{Machine Learning Techniques for Resource Management}
	        \centering
	        \begin{tabular}{|p{3.4cm} | p{3.5cm} |p{3.7cm} |p{6.3cm}|}
	        \hline
	        Learning techniques& Key characteristic& Examples& Usages of graphical models \\
	        \hline
          \vspace{0.18em} Reinforcement learning& \vspace{0.18em}Govern agents in an environment to maximize cumulative rewards &\vspace{0.18em} Opportunistic access \cite{alnwaimi2015dynamic}\cite{noroozoliaee2013efficient}
           Joint power control and relay selection \cite{naddafzadeh2010distributed}
            & \begin{itemize}[leftmargin=*]
               \item Information sharing over edges 
                \item Environment modeling based on graphs 
               \item Performance analysis based on graph theory 
           \end{itemize} 
           \\
	        \hline
	     \vspace{0.18em}Deep learning &\vspace{0.18em} Use a cascade of layers of nonlinear processing units for feature extraction and transformation \vspace{1.2em}&\vspace{0.18em} Algorithm approximation \cite{sun2017learning} Environment/ policy modeling \cite{mnih2015human} \cite{mao2016resource}& \begin{itemize}[leftmargin=*]
	         \item Neural network sparsification based on graphs to simplify the learning procedure
	     \end{itemize}\\
	     
	        \hline
    \vspace{0.18em}Semi-supervised learning &\vspace{0.18em} Infer information of unlabeled data from a small amount of labeled data\vspace{1.2em} &\vspace{0.18em}Vertex labeling \cite{ravi2016large} &\begin{itemize}[leftmargin=*]
        \item Propagating node information over edges 
    \end{itemize}  \\
	         \hline
	        \end{tabular}
	        \label{table:t1}
	\end{table*}
\begin{figure}
    \centering
    \includegraphics[width=0.4\textwidth]{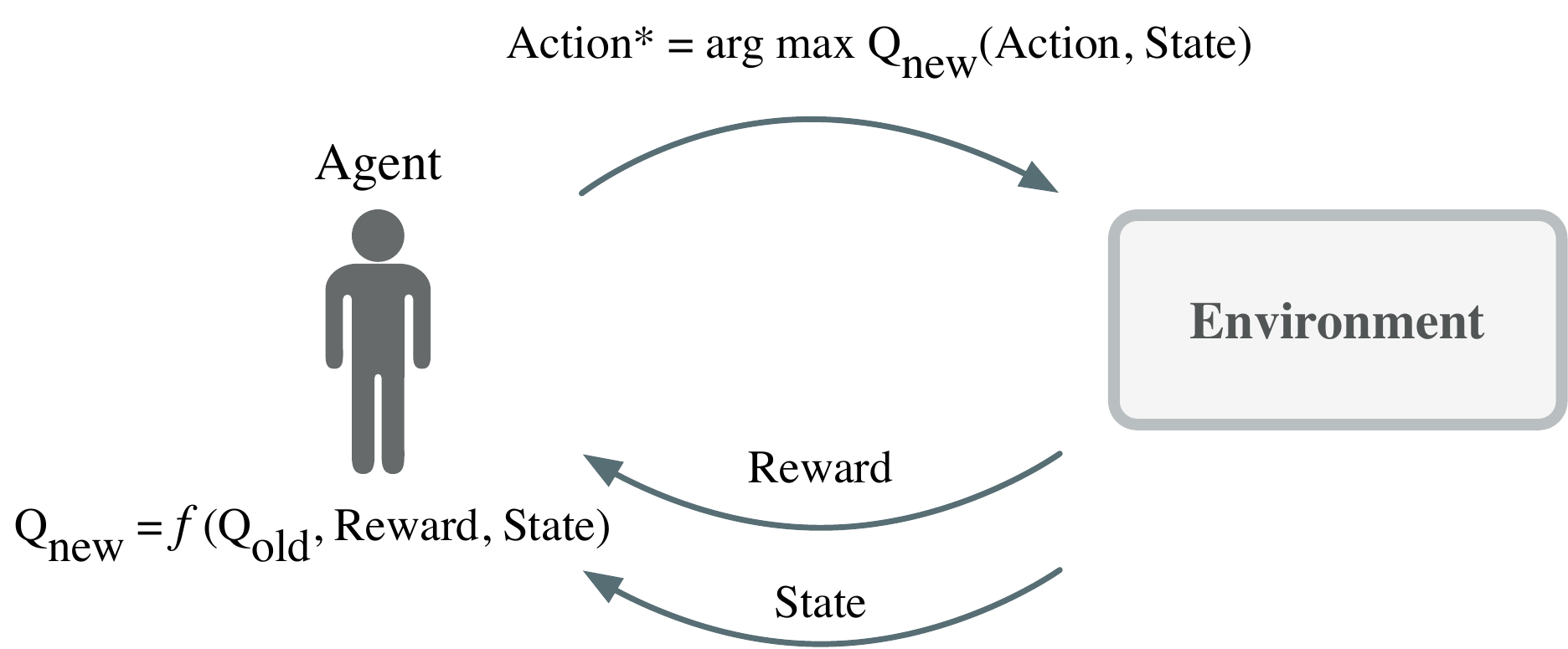}
    \caption{In Q-learning, the Q-function is iteratively updated through observing the actual reward after the agent carries out an action in a given environment.}
    \label{fig:my_label}
\end{figure}
\subsection{Reinforcement Learning}
Reinforcement learning is concerned with how agents ought to take proper actions in a dynamic environment so as to maximize certain cumulative rewards. Regarding network entities as agents and the network as the environment, reinforcement learning is an ideal tool for resource management in H-UDNs. Here, we take a model-free reinforcement learning technique, say Q-learning, as an example. As shown in Fig. 5, Q-learning is used to find an optimal action-selection policy by maximizing the accumulated reward, which is approximated by a Q-function. The Q-function is iteratively updated through observing the actual reward after the agent carries out an action in a given environment. For instance, \cite{alnwaimi2015dynamic} proposed a Q-learning algorithm for femtocells to design transmission schemes based on their own actions and received reward signals in order to maximize the reward functions. In \cite{alnwaimi2015dynamic}, femtocells act as a secondary network and perform opportunistic channel access in a distributed manner, which reduces the implementation complexity. One key challenge of applying existing Q-learning algorithms to H-UDNs is that when the devices aim at maximizing their own reward functions, their behaviors often reduce each other's rewards. This is because a H-UDN consists of various kinds of devices that may have conflicting interests. To address the challenge, it is necessary to introduce a certain level of information exchange among different devices. For example, \cite{naddafzadeh2010distributed} proposed to replace individual reward functions by a global reward function to enhance the performance of joint power control and relay selection. However, obtaining the global reward requires collecting all the reward signals throughout the whole network, which is impractical due to the large network size. Hence, an approximation approach is proposed in \cite{naddafzadeh2010distributed} based on the coverage graph. That is, each device approximates the global reward by averaging the rewards received from the neighboring devices in the coverage graph. Another way to avoid collecting the reward signals from the entire network is to design the scheme in a way that a good system-level performance is achieved when individual reward functions are optimized. \textcolor{black}{Ref. \cite{noroozoliaee2013efficient} considered an opportunistic spectrum access (OSA) problem in cognitive radio networks, where individual objective functions are derived for each user by decoupling the global reward function. Specifically, the OSA problem is transferred to a coloring problem over a conflict graph of the network with users being the vertices. For each user, the individual objective function consists of its own rewards and the rewards of users conflicting with it. It is shown that the proposed algorithm can achieve a near-optimal performance when the conflict graph is a complete graph.}

\subsection{Deep Learning}
Deep learning is potentially a useful tool for resource management in H-UDNs due to its capability of modeling complex non-linear relationships. Deep neural networks can be used to replace the existing high-complexity resource allocation algorithms. Recently, Sun \textit{et al.} approximated weighted MMSE (WMMSE) based power allocation using deep learning by treating the input and output of WMMSE as an unknown non-linear mapping \cite{sun2017learning}. Unlike \cite{borgerding2016onsager}, which approximates the AMP algorithm by mimicing each iteration by a layer of the neural network, \cite{sun2017learning} used a fully-connected neural network to approximate the WMMSE algorithm (see Fig. 4(b)). Through simulations, the authors showed that the fully-connected deep neural network can achieve similar performance as that of WMMSE with much lower complexity. Recall that the power allocation of each transmitter only depends on the power allocation of nearby transmitters. Hence, sparsifying the proposed neural network based on the conflict graph has a great potential to improve the training efficiency without causing noticeable performance loss. 
\par 
Neural networks can naturally be applied to learn the network environments, such as prior information, reward functions, states of agents, and interactions between agents, which are useful in reinforcement learning schemes. This leads to deep reinforcement learning algorithms. In \cite{mnih2015human}, the authors proposed to use a deep convolutional neural network to approximate the optimal action-value function of Q-learning. In this way, successful policies are directly learned from high-dimensional inputs in the end-to-end system with very little model information. In \cite{mao2016resource}, a deep reinforcement learning algorithm is proposed for job scheduling with policy represented by deep neural networks. Through simulations, it is shown that the proposed algorithm performs comparably or even better than standard heuristic algorithms. Moreover, it does not require any prior information of the system environment. \textcolor{black}{Recently, Wang \textit{et al.} proposed to replace traditional DNNs in deep reinforcement learning algorithms with a graph neural network, constructed based on the graphical representation of actions \cite{wang2018nervenet}. The proposed algorithm not only achieves comparable performance to state-of-the-art reinforcement learning methods, but also shows outstanding capability of generalization to different scenarios.} These encouraging observations indicate the great potential of developing deep reinforcement learning techniques for resource management in H-UDNs. Exploiting the graphical representation of underlying networks to improve the efficiency is undoubtedly a subject worth pursuing.
\subsection{Semi-Supervised Learning}
Semi-supervised learning is a powerful tool to train prediction systems with little supervision (i.e., only a small percentage of the training data is labeled). The learning procedure is usually performed over a graph, in which a small fraction of vertices are labeled. The vertices in the partially-labeled graph are connected by an edge if they share some similarities, with the edge weight quantifying the similarities. Then, the learning algorithm gradually labels the unlabeled vertices by propagating information across the graph in a distributed manner, leading to efficient implementation on large graphs \cite{ravi2016large}. Many resource allocation problems in H-UDNs can be interpreted as vertex labeling (a.k.a., graph coloring) problems. For example, frequency channels can be allocated by labeling the devices with different labels (say, different subchannels). In practice, the subchannels occupied by devices at the network boundaries are predetermined to avoid collisions with nearby networks, which means a small number of vertices corresponding to boundary devices are pre-labeled. The radio resource allocation problem thereby can be transferred to a semi-supervised learning problem by carefully defining the edge weights (i.e., similarities). For example, we can set the weights between devices to be inversely proportional to the distances between them to avoid interference among nearby devices. Another example is content placement in network caching. Vertices corresponding to caches are connected in a cooperation graph, only when they are capable of jointly serving a user with the same contents. Constraints in terms of storage capacity, fronthual capacity, transmission delay, etc., can be treated as pre-labeled data in semi-supervised learning algorithms.

\section{Conclusions and Discussions}
\textcolor{black}{This article discussed machine learning as a promising solution for collaborative signal detection and resource management in H-UDNs with graphical representations. Incorporating the network structure by means of various graph representations, machine learning algorithms exhibit great potential in dealing with large-scale systems without relying on mathematically convenient models.} The few examples discussed in this paper serve as an inspiration that calls for future research efforts in this field. \textcolor{black}{There is no doubt that machine learning algorithms associated with graphical representations, once successfully designed, will significantly improve the performance of H-UDNs and consequently bring enormous economic benefits to mobile networks.}

Some potential challenges of applying machine learning in H-UDNs are listed below. First, as discussed before, how to incorporate various prior information in learning algorithms is an interesting future research topic. Meanwhile, the cost of acquiring prior information justifies a thorough investigation on the tradeoff between algorithm performance and the amount and type of prior information acquired (e.g., the exact distribution of a wireless channel vs. the family of distributions of a wireless channel). Secondly, the scalability in terms of processing cost, memory requirements, computation time, etc., is a key factor that determines the feasibility of machine learning algorithms in large-scale H-UDNs. It is therefore worthy to investigate the design of scalable machine learning algorithms by exploiting the underlying structure of physical networks. Thirdly, the network environment of H-UDNs varies dynamically due to the dynamics of mobile users, wireless channels, and interaction between network entities. A good machine learning algorithm must be able to adapt to environment dynamics in a timely and stable manner. 

\bibliographystyle{IEEEtran}
\bibliography{database}

\end{document}